\documentclass[fleqn,10pt]{wlscirep}
\usepackage{amsmath}
\usepackage{amssymb}
\usepackage{graphicx}% Include figure files
\usepackage[utf8]{inputenc} 
\usepackage{floatrow}  
\usepackage{color}
\usepackage{url}
\usepackage{colortbl}
\definecolor{mygray}{gray}{.8}

\title{Rapid online solid-state battery diagnostics with optically pumped magnetometers}

\author[1,2]{Yinan~Hu\footnote[1]{Corresponding author; Electronic mail: yinanhu1@uni-mainz.de}}
\author[1]{Geoffrey~Z.~Iwata}
\author[1]{Lykourgos~Bougas}
\author[2]{John~W.~Blanchard}
\author[1,2]{Arne~Wickenbrock}
\author[1]{Gerhard~Jakob}
\author[4]{Stephan~Schwarz }
\author[4]{Clemens~Schwarzinger}
\author[3]{Alexej~Jerschow}
\author[1,2,5]{Dmitry~Budker}

\affil[1]{Johannes Gutenberg-Universit{\"a}t  Mainz, 55128 Mainz, Germany}
\affil[2]{Helmholtz-Institut Mainz, GSI Helmholtzzentrum f{\"u}r Schwerionenforschung, 55128 Mainz, Germany}
\affil[3]{Department of Chemistry, New York University, New York, NY 10003, USA}
\affil[4]{Institute for Chemical Technology of Organic Materials, Johannes Kepler
University Linz, 4040 Linz, Austria}
\affil[5]{Department of Physics, University of California, Berkeley, CA 94720-7300, USA}
\begin{abstract}
Solid state battery technology is motivated by the desire to deliver flexible power storage in a safe and efficient manner. The increasingly widespread use of batteries from mass-production facilities highlights the need for a rapid and sensitive diagnostic for identifying battery defects. We demonstrate the use of atomic magnetometry to measure the magnetic fields around miniature solid-state battery cells. These fields encode information about battery manufacturing defects, state of charge, impurities, or can provide important insights into ageing processes. Compared with SQUID-based magnetometry, the availability of atomic magnetometers, however, highlights the possibility for a low-cost, portable, and flexible implementation of battery quality-control and characterization technology. 
\end{abstract}

\begin{document}
\flushbottom
\maketitle
\thispagestyle{empty}

\section{Introduction and Background}
%\subsection{Solid state Batteries} 
Lithium-ion rechargeable batteries (LIBs) currently are considered the most promising secondary battery technology to power mobile devices, electric transportation, and power tools. Conventional LIBs typically employ liquid or gel-based electrolytes. These are often volatile and flammable, which gives rise to safety concerns \cite{Li2018,goodenough2010}. Solid-state batteries incorporate ion conducting solid electrolytes instead, which avoid the issue of volatility and leakage. Among other things, solid electrolytes allow the use of electrode materials with high voltages (\textgreater 5\,V), because of their wider electrochemical window.  Furthermore, the use of solid electrolytes could enable the use of lithium metal anodes, thereby providing a viable pathway for higher capacity devices, while maintaining the inherent safety of the battery\cite{Gon2015}. Significant technological challenges need to be overcome to make solid-state-based LIBs a wide-spread reality, including, for example, the facilitation of fast ion transport and the establishment of good interfacial properties between electrodes and the electrolyte medium.

%\subsection{Battery diagnostics tools} 
All these factors highlight the need for the development of advanced diagnostic methodology that would allow proper device characterization at different stages of a battery's life. Currently used techniques which can give in-situ/operando information include ultrasound diagnostics\,\cite{James B2019}, synchrotron-based scanning transmission X-ray microscopy\,\cite{Jongwoo2016}, X-ray micro-diffraction\,\cite{Jun2010}, and Raman spectroscopy\,\cite{Panitz2001}. Manufacturers typically perform electrochemical testing and limited 2D X-ray scanning\,\cite{Vanessa2018,yrynen2012}. X-ray tomography is another powerful (but costly) tool that is generally too slow for high-throughput use with commercial cells\,\cite{Vanessa2018,Yufit2011}. Ultrasound diagnostics are useful for characterization of cells based on changes in density and mechanical properties\,\cite{Hsieh2015,Davies2017}, but do not allow non-contact measurements limiting their broad application.

Rapid techniques that can be used non-destructively are typically limited to electrical impedance spectroscopy, and lately ultrasound techniques. 

Recently, it was also demonstrated that magnetic susceptibility changes within cells could be measured without contact, and non-destructively using an inside-out MRI (ioMRI) technique that used the $^1$H nuclear spin resonance frequencies in water to measure the susceptibility-induced magnetic field changes surrounding a cell when placed in a strong magnetic field\cite{Andrew2018}. It was further shown that the changes in the magnetic susceptibility could be tracked across the charge-discharge cycle, and that these changes followed the expected trends of the lithiation state of the cathode material. This approach therefore provided a single-point state-of-charge measurement and allowed for the identification of inhomogeneities or non-idealities of charge storage in electrochemical cells\,\cite{Andrew2018}. 
Extensions of this method have been developed to encompass the characterization of electrical current distributions (operando) \cite{mohammadi2019}, robust imaging protocols in the presence of strong background magnetic fields or distortions\cite{romanenko2019distortion,romanenko2020accurate}, and a technique for mapping alternating magnetic fields produced by applied AC currents\cite{Stefan2020}. The latter approach is developed with the goal of providing a type of localized electrical impedance spectroscopy. It was shown that ioMRI and an ultrafast inside-out NMR technique can be used to classify cells based on different types of defects\cite{Andrew2018,Roberta2020}. 

%\subsection{Battery diagnostics through magnetic properties} 
Superconducting quantum interference device (SQUID) based magnetic properties measurement systems (MPMS) were used to demonstrate the relation between battery charge state and susceptibility in the lab\,\cite{Gregor2018,Quantum2002,Chernova2011,Stefan2016}. An MPMS is usually used to measure the material susceptibility with high precision, and it contains two main parts: a SQUID magnetometer to measure the induced field from the sample and a superconducting magnet to provide the background field. The MPMS used in this work fits well for material-properties evaluation, but measurements are time consuming and cannot be done with a full-size commercial cell, due to the limitation of the sample size [$\textless (6 \times 6 \times 9$\,mm$^3)$].

\begin{figure}[ht]
\centering
\includegraphics[width=.9\linewidth]{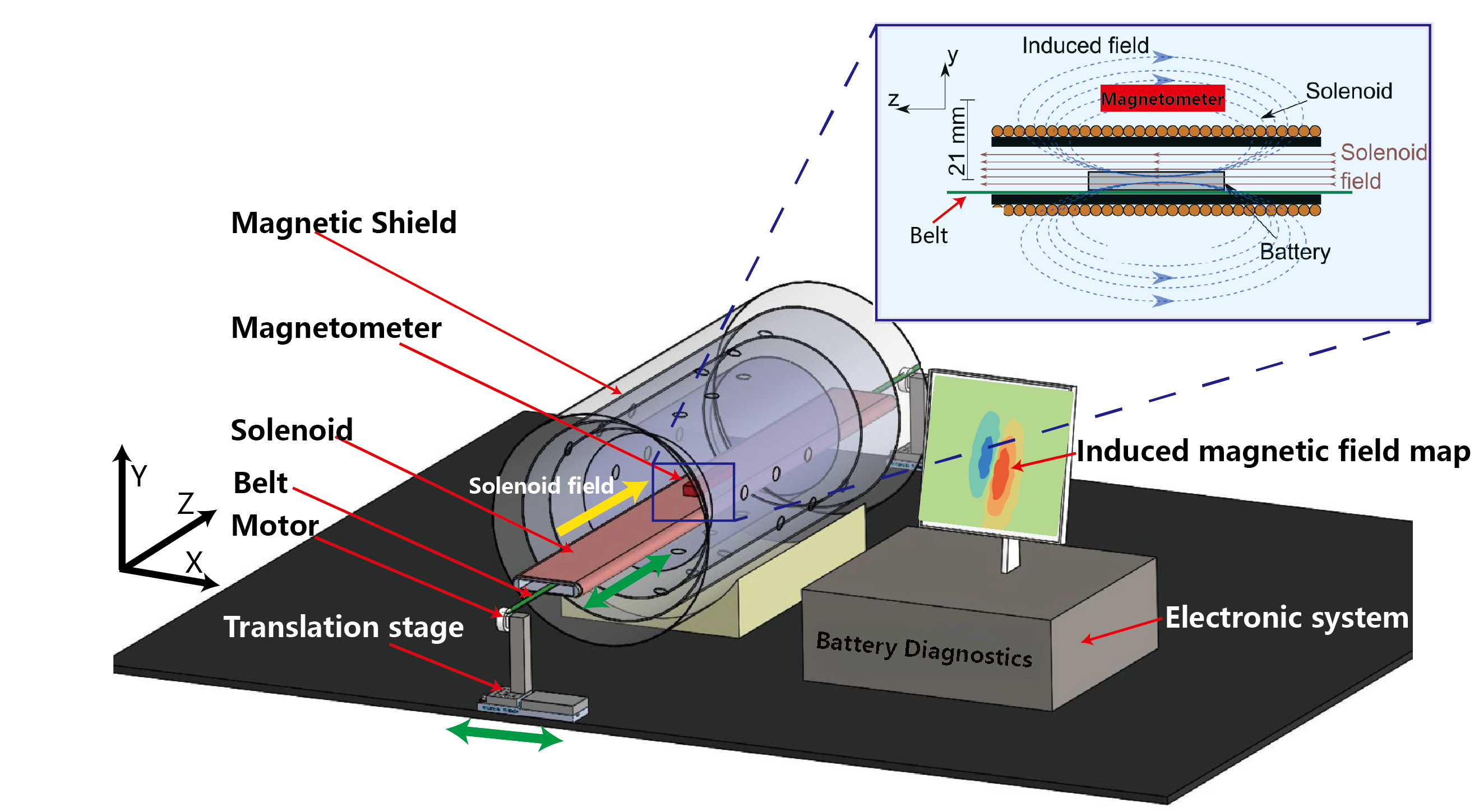}
\caption{Experiment setup.}
\label{fig:Experiment setup}
\end{figure}

Our previous work introduced a diagnostic method\,\cite{hu2019battery} based on atomic magnetometry. The rationale for this approach was the possibility to use these very sensitive sensors (pT$/\sqrt{\rm{Hz}}$ to fT$/\sqrt{\rm{Hz}}$ sensitivity) to report on tiny magnetic field changes around cells as a function of charge state. Tested on Li-ion pouch cells, the measurements showed sensitivity to microampere-level transient internal currents and changes in the magnetic susceptibility of the battery depending on its charge state. The susceptibility measurements involved a long, flat solenoid piercing a magnetic shield, with magnetic field sensors placed outside the solenoid [see Fig.\,\ref{fig:Experiment setup}]. In this configuration, the sensor does not ``see'' the magnetic field of the long solenoid; however, it is fully sensitive to the induced field produced by the battery in the presence of the solenoid field. 

Here, we extend the measurements to miniature commercial solid-state batteries (both operational and defective) and benchmark the atomic-magnetometer results with those based on SQUIDs (Quantum Design MPMS-XL-5). The sensitivity of atomic magnetometers provides opportunities for device characterization to sense small differences in magnetic properties as a function of device history or state. 

For these chip-based cells, in particular, overheating leads to loss of saturation magnetization, and hence is easily identifiable in the measurements. The studies are further supported by Inductively Coupled Plasma Mass Spectrometry (ICP-MS) in order to identify the composition of the cells under study and the origin of the magnetization in these measurements.

\section{Materials and methods}
\subsection{Solid-state battery cells}

\begin{figure}[ht]
\centering
\includegraphics[width=.9\linewidth]{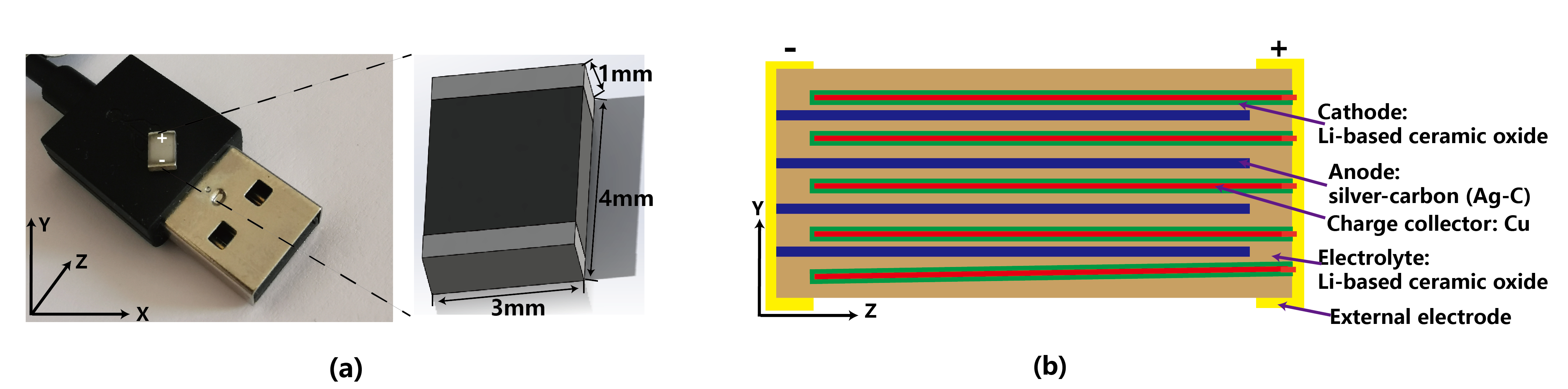}
\caption{(a) Rechargeable solid-state battery (CeraCharge$^{TM}$) from TDK Corporation [www.tdk-electronics.tdk.com]. (b) Z-Y cross-section view of the battery\cite{biohit2018}.}
\label{fig:battery}
\end{figure}

All the measurements were performed on a commercially available all-ceramic solid-state battery (CeraCharge$^{TM}$), which is a rechargeable surface-mount device (SMD) pictured in  Fig.\,\ref{fig:battery} (a). The battery is designed as a chip-scale multi-layer passive component.
It incorporates a Li-based ceramic solid electrolyte. The central copper electrode is used to collect electrons [see Fig.\,\ref{fig:battery} (b)]. The capacity of the battery is 100\,$\mu$Ah at a rated voltage of 1.4\,V.

\subsection{Measurement apparatus}

 The measurements in our setup\cite{hu2019battery} were automated with the use of a ``conveyor-belt'' moving the battery across the measurement region within a second.  The belt moved the battery past the sensors for scanning along the z-coordinate. The cell was automatically translated along the x direction for the next scan while it was transported back to the original z position. Magnetic field maps recorded with the solenoid-current turned on, were consistent with maps expected for a rectangular block with approximately uniform susceptibility (similar to a dipole-field), and the induced field could be calculated precisely. All measurements in this work were carried out at room temperature. 

\subsection{Cell preparation and analyses}

The cells are prepared in different states, which include healthy cells that can power the equipment like the LED normally, broken cells (that were heated to a high enough temperature that the voltage after cool-down is zero, and that can no longer be recharged to a level where they can power equipment), ``healthy'' cells (that can power equipment normally)
demagnetized with a degaussing machine and magnetized with a permanent magnet with a field of 0.05\,T applied for 20\,s. 

The SQUID-magnetometry measurement was carried out with a commercial setup (Quantum Design MPMS-XL-5), and a gelatine capsule was used as a sample holder to position the cell. 

For each ICP-MS measurement, two cells were microwave-digested in 4\,mL HNO$_3$ at 210$^{\circ}$\,C. In order to dissolve further inorganic components, in a second step, the digestion was repeated with 1\,mL HCl at the same conditions. The sample was subsequently diluted to 50\,mL with 18 MOhm water. Low concentrations were measured from this solution and highly concentrated elements after 1:100 dilution with 1\% HNO$_3$. ICP-MS measurements were performed using an XSeries 2, Thermo Scientific ICP-MS instrument.

\begin{figure}[ht]\centering 
\centering
\includegraphics[width=1.0\linewidth]{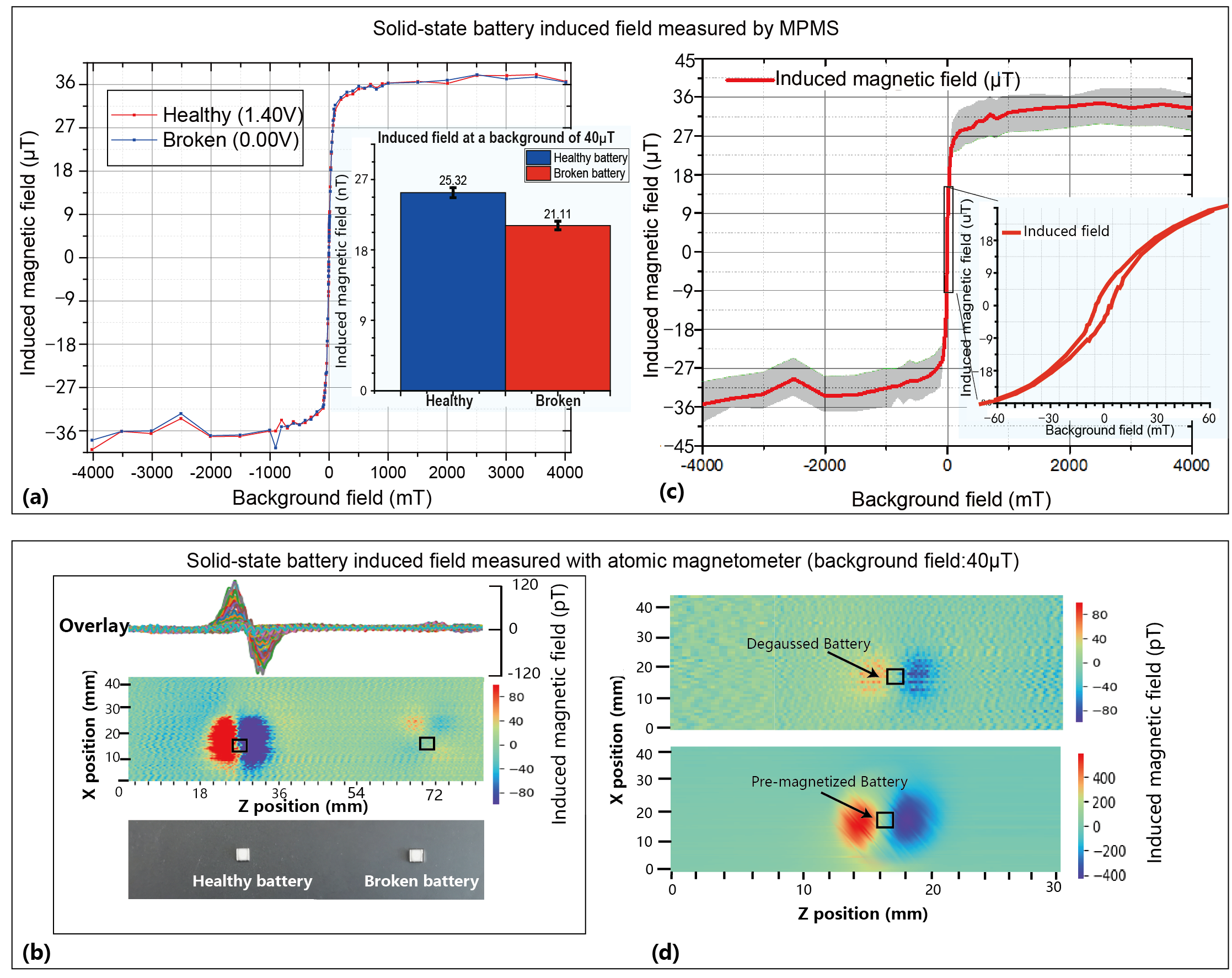}
\caption{(a) Induced magnetic field of the healthy cell and broken cell measured with MPMS. (b) Induced magnetic field vs. background magnetic field as measured by the SQUID-based MPMS.  (c) Induced magnetic field map of the healthy cell and broken cell. (d) Induced magnetic field map of the cell after degaussing and magnetizing.}
\label{fig:statis2}
\end{figure}

\section{Results and Discussion}

The MPMS measurements show that the cells under study here produced stronger induced magnetic field than the cells studied in the earlier work\cite{hu2019battery}. In addition to paramagnetic behavior, a dominating  ferromagnetic behavior is observed [see Fig.\,\ref{fig:statis2} (c),the grey bar is the variation]. From the B-H curve, we can infer that our system behaves predominantly as a soft ferromagnetic material but contains a hysteresis loop in the low-field region (see inset).

An examination of ICP-MS results [see Table.\,\ref{table:icp}] indicates that a possible origin of the ferrormagnetic components is nickel metal. While nickel can also be a component of cathode materials, for typical cathodes, the amount identified is much lower than what would be needed to account for our observations, for example for different types of  Nickel Manganese Cobalt Oxide~(NMC) materials\cite{kaboli2020}. The assumption that the saturation magnetization determined from SQUID magnetometry is due to metallic nickel is supported by the amount of nickel measured by the ICP-MS method. The saturation magnetic field for nickel is 58.57$\pm$0.03 emu$/$g\cite{danan1968new}, the nickel's weight difference between cells is around 0.99$\times10^{-3}$\ mg, therefore we obtain the saturation magnetic field difference as approximately 0.6$\times10^{-4}$\ emu ( $=$6.19$\mu$T ), which fits the experiment results very well [see Fig.\,\ref{fig:statis2} (c)].  

\begin{table}
\small
\renewcommand\arraystretch{1.4}
\centering
\caption{Elemental content of the cell measured with Inductively Coupled Plasma Mass Spectrometry. The mass of the cell is 0.045g. Mn, Fe, and Co could not be detected above the detection threshold. }
\begin{tabular}{cccccc}   
\toprule
 element &&unit && value \\  
\midrule  
  Li  && mg/kg   && 12685 $\pm$ 500 \\
  Mg  && mg/kg   && 4 \\
  Al  && mg/kg   && 417 $\pm$ 60 \\
  K  && mg/kg   && 103$\pm$  9 \\
  Ca  && mg/kg   && 24 \\
  V  && mg/kg   && 27889 $\pm$ 2000\\
  Cr  && mg/kg  && 42 \\
  Mn  && mg/kg  && --- \\
  Fe  && mg/kg  && --- \\
  Co && mg/kg  && --- \\
 \rowcolor{mygray}
  Ni && mg/kg  && 240 $\pm$ 22 \\
  Cu && mg/kg  && 188628$\pm$ 9000 \\
  Zn && mg/kg  && 271 \\
  Pb && mg/kg  && 6 \\
\bottomrule
\end{tabular}
\label{table:icp}
\end{table}

%The saturation magnetization also shows variation from cell to cell (variation range within 8.2$\mu$T)[see Fig.\,\ref{fig:statis2}(c)]. This is likely caused by slight differences in the amount of nickel in the cells.  

%\textbf{[** I think it is better to switch the left and right panels in Fig. 3, also perhaps best to label each pane a, b, c, d, makes it easier to refer to. Below, I am writing the text as if they are switched **]}

Figure 3a shows MPMS measurements for a healthy cell and one that was overheated above its highest working temperature of 80$^{\circ}$\,C. As seen here, the B-H curves are indistinguishable. The inset shows the induced magnetic field values extracted from these curves by a linear fit in the field region above and below $\pm1$\,T. 

Figure 3b shows the comparison of the measurements performed using the atomic magnetometer. For  40 different measured cells (20 healthy and 20 broken ones), the broken cells always showed a different magnetic field pattern, and induced field of broken batteries was weaker; the upper plot in this figure shows the overlaid data, the middle plot is the induced field map, the bottom image shows that healthy and broken cells do not differ in their outward appearance. 

In order to provide an independent verification for the origin of the magnetization that gives rise to the measurements, we performed these measurements by first pre-magnetizing the cell with a 0.05\,T magnet and comparing the measurement with one from a cell that has undergone degaussing. The results show that the pre-magnetized cells show similar values as the healthy cells measured, and the degaussed cells show significantly reduced magnetization, similar to the broken (heated) cell[see Fig.\,\ref{fig:statis2}~(d)]. 

%The overall measured magnetic field values correspond well with the values expected based on the measured amounts of nickel. 

As the results show, our battery scanner could rapidly measure that the magnetization becomes weaker or stronger, depending on whether the cells are either degaussed or magnetized. In particular, these measurements allow one to immediately conclude that a cell has been overheated, for example, something that is of concern for chip-based devices.

MPMS is an ideal tool to study magnetic properties of materials, and is widely applied to the battery prototype research\cite{Stefan2016}. It can evaluate the charge-state of the battery precisely by monitoring the transition metal's magnetic property during the charge or discharge period\cite{klinser2016continuous,Chernova2011}. In this work, MPMS could provide a high background magnetic field (-4 to 4\,tesla) providing access to the complete B-H curve, while our piercing solenoid system could only provide a stable background magnetic field from -80 to 80\,$\mu$T. However, the resolution of the MPMS background field is 33\,$\mu$T, the resolution of our setup background field is 30nT. Our system does not have the residual magnetization problem, while the MPMS leaves a residual magnetic field of $\approx500\,\mu$T. Finally, the atomic magnetometer system had a sub-pT sensitivity, while the  MPMS sensitivity was 1$\times10^{-8}$\ emu, corresponding to around 12.6\,pT. Thus our system is better suited for measurements at low fields, in order to identify ferromagnetic contributions influenced by temperature\cite{coey2010magnetism}[see Fig.\,\ref{fig:statis2} (b)].

Furthermore, the atomic magnetometer setup could produce a diagnostic map of the battery cells' induced magnetic field within 20\,s, providing resolution of the distribution of magnetic particles. A high-speed motor or a magnetometer array instead of a single sensor could even be used for these non-contact diagnostics of batteries in a sub-second time scale. These measurements can be performed in-situ and operando with different-sized cells and under factory conditions. We show batteries damaged by heating can be identified; this is particularly important for surface-mounted devices. 

The magnetic field detection sensitivity of our setup is better than 400 fT$/\sqrt{\rm{Hz}}$ from 1 to 150Hz, which could help to detect a milligram difference in nickel content between cells. The dynamic range of our atomic magnetometer is approximately 5 nT, which also puts an upper limit on the nickel concentration. Recent developments in earth-field atomic magnetometers with larger dynamic ranges \cite{limes2020portable} could solve this problem.
%[** here we should include a paragraph about sensitivity and detection sensitivity of magnetic components, e.g. what would be the detection limit of Ni? **]
These direct magnetic field measurements with atomic magnetometers provide hitherto inaccessible opportunities for cell characterization, classification, and monitoring in research and industry. Apart from battery diagnostics, the technique can be applied to general magnetic-susceptibility based diagnostics, as well as for residual magnetization measurements, for example, in rock-magnetism studies and magnetic materials testing\cite{dang2010}.

\section{Conclusion} 

We have examined here the possibility of using atomic magnetometers to characterize the magnetic properties of rechargeable solid-state Li-ion batteries, in a rapid, non-destructive and non-contact fashion. The measurements are compared to SQUID magnetometry based measurements and supported by an elemental analysis. The magnetic components (nickel) in the cells are determined to be the source of the measured magnetization, and atomic magnetometry provides a convenient means of determining whether cells have been overheated. In a broader sense, atomic magnetometry provides sufficient sensitivity to sense the dependence of the magnetic susceptibility on the charge state and to identify certain defects or cell history. Given the difficulty of characterizing solid-state cells, such non-destructive and non-contact techniques could become useful tools in the analysis of cells at different stages of their life cycle.

\section*{Acknowledgement} 

The work was funded in part by a grant by the U.S. National Science Foundation under award CBET 1804723 and the German Federal Ministry of Education and Research (BMBF) within the Quantumtechnologien program (FKZ13N14439).

\newpage

\renewcommand{\bibname}{\bfont References}
\bibliographystyle{elsarticle-num}

\begin{thebibliography}{99}

\bibitem{Li2018}
  Li Matthew,
  30 years of lithium-ion batteries,
  Advanced Materials,
  volume {30},
  number {33},
  pages {1800561},
  {2018}

\bibitem{goodenough2010}
  Goodenough John B and Kim Youngsik,
  Challenges for rechargeable Li batteries,
  Chemistry of Materials,
  volume {22},
  number {3},
  pages {587--603},
  {2010}
 
\bibitem{Gon2015}
  Kim, Joo Gon,
  A review of lithium and non-lithium based solid state batteries,
  Journal of Power Sources,
  volume {282},
  pages {299--322},
  {2015}


\bibitem{James B2019}
  Robinson, James B,
  Spatially resolved ultrasound diagnostics of Li-ion battery electrodes,
  Physical Chemistry Chemical Physics,
  volume {21},
  number {12},
  pages {6354--6361},
  {2019}

  
\bibitem{Jongwoo2016}
  Lim Jongwoo,
  Origin and hysteresis of lithium compositional spatiodynamics within battery primary particles,
  Science,
  volume {353},
  number {6299},
  pages {566--571},
  {2016}

\bibitem{Jun2010}
  Liu Jun,
  Visualization of Charge Distribution in a Lithium-Ion Battery Electrode,
  Meeting Abstracts,
  number {4},
  pages {208--208},
  {2010}


\bibitem{Panitz2001}
  Panitz, Jan-Christoph,
  Raman microscopy applied to rechargeable lithium-ion cells-Steps towards in situ Raman imaging with increased optical efficiency,
  Applied Spectroscopy,
  volume {55},
  number {9},
  pages {1131--1137},
  {2001}


\bibitem{Vanessa2018}
  Wood, Vanessa,
  X-ray tomography for battery research and development,
  Nature Reviews Materials,
  volume {3},
  number {9},
  pages {293--295},
  {2018}



\bibitem{yrynen2012}
  V{\"a}yrynen, Antti,
  Lithium ion battery production,
  The Journal of Chemical Thermodynamics,
  volume {46},
  pages {80--85},
  {2012}



\bibitem{Yufit2011}
  Yufit V,
  Investigation of lithium-ion polymer battery cell failure using X-ray computed tomography,
  Electrochemistry Communications,
  volume {13},
  number {6},
  pages {608--610},
  {2011}



\bibitem{Hsieh2015}
  Hsieh AG,
  Electrochemical-acoustic time of flight: in operando correlation of physical dynamics with battery charge and health,
  Energy \& Environmental Science,
  volume {8},
  number {5},
  pages {1569--1577},
  {2015}



\bibitem{Davies2017}
  Davies, Greg,
  State of charge and state of health estimation using electrochemical acoustic time of flight analysis,
  Journal of The Electrochemical Society,
  volume {164},
  number {12},
  pages {A2746--A2755},
  {2017}



\bibitem{Andrew2018}
  Ilott, Andrew J,
  Rechargeable lithium-ion cell state of charge and defect detection by in-situ inside-out magnetic resonance imaging,
  Nature Communications,
  volume {9},
  number {1},
  pages {1--7},
  {2018}

\bibitem{mohammadi2019}
  Mohammadi Mohaddese and Silletta,
  Diagnosing current distributions in batteries with magnetic resonance imaging,
  Journal of Magnetic Resonance,
  volume {309},
  pages {106601},
  {2019}
  
\bibitem{romanenko2019distortion}
  Romanenko Konstantin and Jerschow Alexej,
  Distortion-free inside-out imaging for rapid diagnostics of rechargeable Li-ion cells,
  Proceedings of the National Academy of Sciences,
  volume {116},
  pages {18783--18789},
  {2019}

\bibitem{romanenko2020accurate}
  Romanenko Konstantin and Kuchel Philip W and Jerschow Alexej,
  Accurate visualization of operating commercial batteries using specialized magnetic resonance imaging with magnetic field sensing,
  Chemistry of Materials,
  volume {32},
  pages {2107--2113},
  {2020}
  

\bibitem{Stefan2020}
Stefan Benders, Mohaddese Mohammadi,
  Mapping oscillating magnetic fields around rechargeable batteries,
  Journal of Magnetic Resonance,
  volume {319},
  number {39},
  pages {1074},
  {2020}

\bibitem{Roberta2020}
Roberta Pigliapochi,  Stefan Benders,
  Ultrafast inside‐out NMR assessment of Rechargeable Cells,
  Batteries \& Supercaps,
  volume {31},
  number {31},
  pages {1022},
  {2020}
  
\bibitem{Alexej2016}
Ilott, Andrew J,
  Real-time 3D imaging of microstructure growth in battery cells using indirect MRI,
  Proceedings of the National Academy of Sciences,
  volume {113},
  number {39},
  pages {10779--10784},
  {2016}

  

\bibitem{Chernova2011}
  Chernova, Natasha A,
  What can we learn about battery materials from their magnetic properties?,
  Journal of Materials Chemistry,
  volume {21},
  number {27},
  pages {9865--9875},
  {2011}



\bibitem{Gregor2018}
  Klinser Gregor,
  Process Monitoring of Charging/Discharging of Lithium Ion Battery Cathodes by Operando SQUID Magnetometry,
  Encyclopedia of Interfacial Chemistry: Surface Science and Electrochemistry: Reference Module in Chemistry, Molecular Sciences and Chemical Engineering,
  pages {849--855},
  {2018}

  
\bibitem{Quantum2002}
  Quantum Design, Inc., San Diego,
  MPMS Application Note 1014-213,
  pages {10--20},
  {04 February 2002}


\bibitem{Stefan2016}
  Topolovec Stefan,
  Charging of lithium cobalt oxide battery cathodes studied by means of magnetometry,
  Solid State Ionics,
  volume {293},
  pages {64--71},
  {2016}

\bibitem{biohit2018}
  Rechargeable solid-state SMD battery for IoT applications,
  \url{www.tdk-electronics.tdk.com/en/2471330/tech-library/articles/applications---cases---video/rechargeable-solid-state-smd-battery-for-iot-applications/2431020},
  {2018}

  
\bibitem{hu2019battery}
  Hu Yinan and Iwata Geoffrey Z,
  Sensitive magnetometry reveals inhomogeneities in charge storage and weak transient internal currents in Li-ion cells,
  Proceedings of the National Academy of Sciences,
  volume {117},
  pages {10667--10672},
  {2019}

\bibitem{kaboli2020}
  Kaboli Shirin and Demers Hendrix and Paolella Andrea and Darwiche,
  Behavior of Solid Electrolyte in Li-Polymer Battery with NMC Cathode via in-Situ Scanning Electron Microscopy,
  Nano Letters,
  volume {20},
  number {3},
  pages {1607--1613},
  {2020},
  
\bibitem{danan1968new}
  Danan H and Herr,
  New determinations of the saturation magnetization of nickel and iron,
  Journal of Applied Physics,
  volume {39},
  pages {669--670},
  {1968}
  
\bibitem{limes2020portable}
  Limes ME and Foley,
  Portable magnetometry for detection of biomagnetism in ambient environments,
  Physical Review Applied,
  volume {14},
  pages {011002},
  {2020}
  
\bibitem{buchner2018tutorial}
  Buchner M,
  Tutorial: Basic principles, limits of detection, and pitfalls of highly sensitive SQUID magnetometry for nanomagnetism and spintronics,
  Journal of Applied Physics,
  volume {124},
  number {16},
  pages {161101},
  {2018}

\bibitem{coey2010magnetism}
  Michael Coey,
  Magnetism and magnetic materials,
  Cambridge university press,
  pages {128-194},
  2010

\bibitem{klinser2016continuous}
  Klinser Gregor and Topolovec Stefan,
  Continuous monitoring of the bulk oxidation states in Li x Ni 1/3 Mn 1/3 Co 1/3 O 2 during charging and discharging,
  Applied Physics Letters,
  volume {109},
  number {21},
  pages {213901},
  {2016}

\bibitem{dang2010}
  Dang HB and Maloof Adam C and Romalis Michael V,
  Ultrahigh sensitivity magnetic field and magnetization measurements with an atomic magnetometer,
  Applied Physics Letters,
  volume {97},
  number {15},
  pages {151110},
  {2010}




\end{thebibliography}

\end{document}